\documentclass[10pt,prd,twocolumn,showpacs,preprintnumbers,amsmath,amssymb,nofootinbib,APS]{revtex4-1}

\usepackage{dcolumn}
\usepackage{bm}
\usepackage[latin1]{inputenc}
\usepackage[spanish,english]{babel}
\usepackage{amsfonts}
\usepackage{amssymb}
\usepackage{graphicx}
\usepackage{color}
\usepackage[normalem]{ulem}

\newcommand{\be}{\begin{equation}}
\newcommand{\ee}{\end{equation}}
\newcommand{\bea}{\begin{eqnarray}}
\newcommand{\eea}{\end{eqnarray}}

\begin{document}
%\preprint{hep-th/}
\title{{\bf Adiabatic regularization and particle creation for spin one-half fields}}

\author{Aitor Landete, José Navarro-Salas and Francisco Torrentí}
\affiliation{ {\footnotesize Departamento de Fisica Teórica and
IFIC, Centro Mixto Universidad de Valencia-CSIC, Facultad de Física, Universidad de Valencia,
        Burjassot-46100, Valencia, Spain. \ \ }}
%\author{Jose Navarro-Salas}
%\affiliation{ {\footnotesize Departamento de Fisica Teorica and
%IFIC, Centro Mixto Universidad de Valencia-CSIC,
 %   Facultad de Fisica, Universidad de Valencia,
  %      Burjassot-46100, Valencia, Spain. \ \ }}
%                \author{Francisco Torrenti}
%\affiliation{ {\footnotesize Departamento de Fisica Teorica and
%IFIC, Centro Mixto Universidad de Valencia-CSIC, Facultad de Fisica, %Universidad de Valencia,
  %      Burjassot-46100, Valencia, Spain.}}

\date{\today}

\begin{abstract}

The extension of the adiabatic regularization method to spin-$1/2$ fields requires  a self-consistent adiabatic expansion of the field modes. We provide here the details of such  expansion, which differs from the WKB ansatz that works well for scalars, to firmly establish the generalization of the adiabatic renormalization scheme to spin-$1/2$ fields. 
%We also provide a general overview of the adiabatic method to analyze particle creation and perform renormalization of relevant expectation values. 
We focus on the computation of particle production in de Sitter spacetime and obtain an analytic expression of the  renormalized stress-energy tensor for Dirac fermions.

\end{abstract} 
 

\maketitle

%%%%%%%%%%%%%%%%%%%%%%%%%%%%%%%%%%%%%

\section{Introduction}\label{Introduction}

Renormalization in curved spacetime is historically tied to the discovery of particle creation in a time-dependent gravitational field \cite{parker66, parker69, parker-toms, birrell-davies}. If the particle number of created particles in an expanding universe is calculated in an assumed  asymptotically  Minkowskian region, the result is unambiguous and finite. However, if the  particle number  operator is evaluated during the expansion, the result has potential ultraviolet divergences (UV) even for a very slow expansion. Adiabatic  regularization was  originally introduced as a way to overcome these UV divergences and the rapid oscillations of the particle number operator \cite{parker66}. The method was later  generalized to consistently deal with the UV divergences of the stress-energy tensor of scalar fields in homogeneous cosmological backgrounds \cite{parker-fulling74}.  The adiabatic regularization 
method starts with the formal 
expression for the expectation values of the stress-energy tensor $\langle T_{\mu \nu} \rangle$. One then performs a large momentum asymptotic expansion and identify the leading terms giving rise to formal UV divergences in the integration over momenta. These terms are the same for all physical states. Adiabatic renormalization proceeds then by subtracting those leading terms in the large momentum expansion. The resulting momentum integral for  the stress-energy tensor is UV finite. Since the adiabatic subtractions in momentum space give rise directly to a finite  momentum integral, the mechanism of adiabatic subtraction is also acting as a regularization procedure. Hence the name of "adiabatic regularization" to refer to the whole process of "renormalization".

The leading terms in the asymptotic series in momenta should be uniquely identified. This is strictly required since  the  adiabatic subtraction actually involves terms for all momenta, even small ones. To unambiguously characterize the leading terms one needs a physically sound and mathematically well-defined procedure. This is naturally offered by the Liouville or WKB-type asymptotic expansion of the mode functions. This procedure was suggested by the analysis of the particle number operator in expanding universes. The use of the WKB-type expansion for the modes to define particles enforces the physical requirement that the mean particle number is an adiabatic invariant. The covariant notion of adiabatic invariance guaranties the underlying covariance of the subtraction procedure. Moreover, one should subtract only the minimum number of terms necessary to obtain a finite result. This way one keeps as much as possible the form of the original expression for the stress-energy tensor 
 \cite{parker-toms}.

The direct method of adiabatic regularization to remove UV divergences  in Friedmann-Lemaître-Robertson-Walker (FLRW) universes is equivalent to the more conventional subtraction procedure  based on the renormalization of coupling constants in Einstein's equations. The three  type of UV divergences (quartic, quadratic, and logarithmic) in the formal expression of the stress-energy tensor $\langle T_{\mu\nu}\rangle$ would be canceled by counterterms associated to the cosmological constant $ \Lambda g_{\mu\nu}$, the Einstein tensor $G_{\mu\nu}$, and  higher-order terms proportional to $^{(1)}H_{\mu\nu}$ \cite{bunch80}.  These three terms are of adiabatic order zero, two, and four, respectively. The rule of minimal subtraction in the general procedure of adiabatic regularization can therefore be  additionally justified in terms of renormalization of coupling constants.

 An alternative asymptotic expansion to consistently identify the subtraction terms  in a generic spacetime was suggested by DeWitt \cite{dewitt75}, generalizing the Schwinger proper-time formalism. The DeWitt-Schwinger expansion was  armed with the powerful point-splitting technique  {\cite{christensen76} and applied, mainly for scalar fields, to different spacetimes of major physical interest \cite{birrell-davies}. The DeWitt-Schwinger point-splitting method for scalar fields was proved to be equivalent to  adiabatic regularization \cite{Birrell78, anderson-parker}. However, a distinguishing 
 characteristic of adiabatic regularization is its capability to overcome the UV divergences occurring in the particle number operator. Moreover, a major practical advantage of adiabatic regularization is that it is very efficient for numerical calculations \cite{Hu-Parker, Anderson, Anderson2}. It is also potentially important to scrutinize the power spectrum in inflationary cosmology \cite{parker07} and to study implications of quantum gravity at low energies \cite{agullo-ashtekar-nelson}.

 The point-splitting prescription \cite{dewitt75, christensen76} 
can be naturally extended  to  spin-$1/2$  fields \cite{christensen78} and one would expect an analogous extension within the adiabatic subtraction scheme.
However, the WKB template that works for scalar field modes is actually closely related to the Klein-Gordon product, but not to the Dirac product. 
In fact, a self-consistent adiabatic expansion for spin one-half modes has been so far elusive. A solution to this problem has been recently sketched  in \cite{LNT}   and the purpose of this paper is to provide the details of the proposed expansion and to firmly establish the extension of the adiabatic regularization to spin-$1/2$ fields. 

 To properly understand the novelties introduced for spin-$1/2$ fields, we briefly review in Sec. II the adiabatic renormalization method for scalar fields. 
%This will be done in section II, while in  section III we will apply it to analyze particle creation in de Sitter spacetime and its associated stress-energy tensor. 
In Sec. III we describe the proposed adiabatic expansion for the spin-$1/2$ field modes to find the renormalization subtraction terms. In Sec. IV we test the consistency of the extended adiabatic method  by working out the conformal and axial anomalies. We also study fermionic particle creation  in a FLRW spacetime. In Sec. V we study the creation of Dirac particles in de Sitter spacetime, and an analytical expression for the renormalized stress-energy tensor is obtained. Finally, in Sec. VI we summarize our main conclusions. Our conventions follows \cite{parker-toms, birrell-davies} with $\hbar=c=1$.

%%%%%%%%%%%%%%%%%%%%%%%%%%%%%%%%%%%%%%

\section{Adiabatic regularization for scalar fields}

The equation of motion of a scalar field $\phi (\vec{x},t)$ of mass $m$ propagating in a curved background is
\be (\Box + m^2 +\xi R) \phi =0 \label{waveeq}\ee
where $\Box \equiv \nabla^{\mu} \nabla_{\mu}$, $R$ is the  Ricci scalar of the metric and $\xi$ is  the coupling of the field to the curvature. If the field propagates in a spatially flat FLRW universe with metric
 \be ds^2=dt^2- a^2(t) (dx^2 + dy^2 + dz^2) \label{frwmetric} \ ,\ee 
Eq. (\ref{waveeq}) takes the form
\be a^{-3} \partial_t (a^3 \partial_t \phi) - a^{-2} \sum_i \partial_i^2 \phi + (m^2 +\xi R) \phi = 0 \ . \label{gen-eqmov2}\ee
We assume that the field satisfies periodic boundary conditions in a cube of comoving length $L$. In that case, $\phi$ can be expanded in terms of mode functions
\be \phi (\vec{x}, t)= \sum_{\vec{k}} ( A_{\vec{k}}f_{\vec{k}}(\vec{x}, t) +  A^{\dagger}_{\vec{k}}f^*_{\vec{k}}(\vec{x}, t)) \label{phifield}
\ee
where  $k^i= 2\pi n^i/L$ with $n^i$ an integer,  $A^{\dagger}_{\vec{k}}$ and $A_{\vec{k}}$ are creation and annihilation operators and %The continuous limit $L \to \infty$ can be recovered by making at the end of the calculations the substitution
%\be L^{-3} \sum_{{\bf k}} \rightarrow (2 \pi)^{-3} \int d^3 {{ \bf k}} \ . \label{minko-limit}\ee
%We assume that the modes $f_{\vec{k}}$ obey the ansatz
\be f_{\vec{k}} (\vec{x}, t) = \frac{1}{\sqrt{2L^3a^3(t)}} e^{i\vec{k}\vec{x}}h_k(t)  \label{phians} \ee
 [$k \equiv |\vec{k}|$]. $h_k (t)$ is a time-dependent function.  By substituting (\ref{phians}) into (\ref{gen-eqmov2}), we find that it satisfies
\be \frac{d^2 h_k}{dt^2} + (\omega_k^2 + \sigma) h_k=0 \label{hqeq}\ee
where $\omega_k(t)= \sqrt{k^2/a^2(t) + m^2}$ is the frequency of the mode and $\sigma \equiv \left( 6\xi - \frac{3}{4} \right)\dot a^2/a^2 + \left( 6\xi - \frac{3}{2} \right) \ddot a/a$. % \be \label{sigma}\sigma \equiv \left( 6\xi - \frac{3}{4} \right) \frac{\dot a^2}{a^2} + \left( 6\xi - \frac{3}{2} \right) \frac{\ddot a}{a} \ . \ee  
[The dot notation means  differentiation  with respect to time $t$]. We require these modes to be normalized with respect to the Klein-Gordon product $(f_{\vec{k}}, f_{\vec{k}'})= \delta_{\vec{k},\vec{k}'}$. This is equivalent to imposing to $h_k (t)$ the Wronskian-type condition
\be \label{Wronskian} h_k^*\dot h_k - \dot h_k^{*} h_k = -2i \ . \ee
This condition ensures the usual commutation relations for the creation and  annihilation operators.
%\be [A_{\vec{k}}, A_{\vec{k}'}^{\dagger}]= \delta_{\vec{k},\vec{k}'} \ ,   \hspace{0.7cm}  [A_{\vec{k}}, A_{\vec{k}'}]= 0=  [A_{\vec{k}}^{\dagger}, A_{\vec{k}'}^{\dagger}] \ .\ee
Differential Eq. (\ref{hqeq}), together with condition (\ref{Wronskian}), leave us with one unspecified degree of freedom for the function $h_k (t)$, and then for the vacuum state $| 0 \rangle$ defined as $A_{\vec{k}} | 0 \rangle \equiv 0$. 
%If we were in Minkowski spacetime (where $a = 1$ and $\sigma = 0$), the natural solutions  of (\ref{hqeq}) would be
%\be h_k^{M} (t) = \frac{1}{\sqrt{\omega_k}} e^{-i \omega_k t} \label{hkmink} \ee
%with $\omega_k = \sqrt{k^2 + m^2}$. These  are positive-frequency solutions which  allow us to interpret the operators $A_{\vec{k}}^{\dagger}$ and $A_{\vec{k}}$ as creation and annihilation operators of particles. However, for a generic scale factor $a(t)$,  (\ref{hkmink}) is not solution of (\ref{hqeq}) and the ambiguity must be solved in another way.
Adiabatic regularization and the definition of physical particles is based on  a WKB-type expansion for the modes. We can substitute into (\ref{hqeq}) the  ansatz
\be \label{ansatzWKB} h_k(t) = \frac{1}{\sqrt{W_k(t)}}e^{-i\int^t W_k(t')dt'} \ ,  \ee
where $W_k (t)$ is  a time-dependent function. This  ansatz obeys  condition (\ref{Wronskian}). We get the following equation for $W_k (t)$:
\be W_k^2 = \omega_k^2 + \sigma + W_k^{-1/2}\frac{d^2}{dt^2}W_k^{-1/2} \ . \label{WKfunc} \ee
%We can now perform a generalized Liouville or WKB-type asymptotic expansion based on this ansatz. 
$W_k (t)$ can be expanded as an adiabatic series $W_k(t) = \omega^{(0)}(t) + \omega^{(1)}(t)+ \omega^{(2)}(t) + \omega^{(3)}(t) + ...$,
%\be W_k(t) = \omega^{(0)}(t) + \omega^{(1)}(t)+ \omega^{(2)}(t) + \omega^{(3)}(t) + ... \label{sumexp} \ee
where the term $\omega^{(n)}$ has $n$ time derivatives of the scale factor $a(t)$. If we impose the leading term $\omega^{(0)}$ to be the physical redshifted frequency $\omega^{(0)}(t)\equiv \omega(t)\equiv \omega_k(t)= \sqrt{k^2/a^2(t) + m^2}$, 
%\be \omega^{(0)}(t)\equiv \omega(t)\equiv \omega_k(t)= \sqrt{k^2/a^2(t) + m^2} \ , \label{omega0order} \ee
the other terms can be obtained by solving (\ref{WKfunc})  at a given adiabatic order. It is found that $\omega^{(1)} = \omega^{(3)} = 0$,
%\be \label{omega2order}\omega^{(2)}(t)= \frac{\sigma}{2\omega(t)} + \frac{1}{2}\omega^{-1/2}(t)\frac{d^2}{dt^2}\omega^{-1/2}(t) \ee
%and 
%\bea  \label{omega4order} \omega^{(4)}(t) &=&  \frac{1}{4}\omega^{(2)}\omega^{-3/2}(t)\frac{d^2}{dt^2}\omega^{-1/2}(t) -\frac{1}{2}\omega^{-1}(t)(\omega^{(2)}(t))^2  \nonumber \\ &-& \frac{1}{4}\omega^{-1/2}(t)\frac{d^2}{dt^2}\left [ \omega^{-3/2}(t)\omega^{(2)} \right ]\ . \eea
 %The second term of the expansion $\omega^{(2)} (t)$ is given in terms of $a(t)$ as
and $ \omega^{(2)} = \frac{5 m^4 \dot a^2}{8 a^2 \omega^5(t)} - \frac{2 m^2 \dot a^2 + m^2 a \ddot a}{4 a^2 \omega^3(t)} - \frac{(\frac{1}{6} - \xi) R}{2 \omega(t)}$. 
%while the corresponding fourth-order contribution is given in appendix I.
% We will also need the adiabatic expansion of the inverse. It is
% \be W^{-1} = \omega^{-1} + (W^{-1})^{(1)} + (W^{-1})^{(2)} + ... \ee
% where $(W^{-1})^{(1)} = (W^{-1})^{(3)} = 0$,
% \be (W^{-1})^{(2)} = - \omega^{-2} \omega^{(2)}\ee
% and
% \be (W^{-1})^{(4)} = - \omega^{-3} (\omega^{(2)})^2 - \omega^{-2} \omega^{(4)} \ . \ee
% The second term of the expansion is written in terms of the scale factor as
% \be (W^{-1})^{(2)}= \frac{(\frac{1}{6}-\xi)R}{2 \omega^3}+\frac{m^2\dot a^2}{2a^2 \omega^5} + \frac{m^2\ddot a}{4a \omega^5} -\frac{5m^4\dot a^2}{8a^2 \omega^7} \ee
% while the corresponding fourth-order contribution $(W^{-1})^{(4)}$ has 30 terms and is given in appendix I.
This expansion constitutes the basic cornerstone of the adiabatic regularization method. It allows us to define the particle number \cite{parker66, parker69} and also to renormalize local operators by removing their UV divergences, while keeping their covariance \cite{parker-fulling74}. 
%The underlying covariance of the subtraction procedure, based on the  covariant notion of adiabatic invariance, makes it a self-consistent renormalization method. 
%The subtraction terms identified using the limiting case of slow expansion $\dot a \to 0, \ddot a \to 0$, etc, remain the same for an arbitrary (non-adiabatic) expansion factor $a(t)$.  

The particle number in an expanding universe is not a constant of motion, but it is, nevertheless,  an adiabatic invariant. Since the particle number is actually changing while it is being measured, there is always an intrinsic uncertainty in the particle number concept. Therefore, one should expect a fuzzy characterization  of the splitting between positive and negative frequency modes.  However, when the expansion enters into the adiabatic regime, the characterization  is naturally done in terms of the nth-order adiabatic modes $g_{\vec{k}}^{(n)} (\vec{x}, t)$, defined as
\be g^{(n)}_{\vec{k}} \equiv \frac{1}{\sqrt{2L^3a^3(t)}} e^{i\vec{k}\vec{x}} g^{(n)}_k(t) \ee
 with 
\be g_k^{(n)} (t) \equiv \frac{1}{\sqrt{W_k^{(n)} (t)}} e^{-i \int^t W_k^{(n)} (t') dt'} \ ,\label{gen-gsol} \ee
 and $W_k^{(n)} (t) \equiv \omega^{(0)} + \omega^{(1)} + \dots + \omega^{(n)}$. 
 %These modes constitute the adiabatic definition of particles given in the pioneer work \cite{parker66}. We assume this concept of particles from now on. 
  %The annihilation and creation operators for real particles have to be time-dependent, thus reflecting the particle creation process. 
  We expand the $\phi$ field as
\be \phi (\vec{x}, t)=\sum_{\vec{k}}\{ a_{\vec{k}}(t)g^{(n)}_{\vec{k}}(\vec{x}, t) +  a^{\dagger}_{\vec{k}}(t) g^{(n)*}_{\vec{k}}(\vec{x}, t)\}
\ , \ee
where the time-dependent operators $a^{\dagger}_{\vec{k}}(t)$ and $ a_{\vec{k}}(t)$ obey the usual commutation relations.
These operators are related with the time-independent ones $A^{\dagger}_{\vec{k}}$ and  $ A_{\vec{k}}$ by  the Bogolubov transformations $a_{\vec{k}}(t)= \alpha_k^{(n)}(t)A_{\vec{k}} + \beta^{*(n)}_k(t)A^{\dagger}_{-\vec{k}}$.
%\be \label{Bogolubov} a_{\vec{k}}(t)= \alpha_k^{(n)}(t)A_{\vec{k}} + \beta^{*(n)}_k(t)A^{\dagger}_{-\vec{k}} \ . \ee
The time-dependent coefficients $\alpha_k^{(n)}(t)$ and $\beta_k^{(n)} (t)$ can be obtained by writing the exact mode functions
 $h_k(t)$ in terms of the adiabatic modes $g_k^{(n)} (t)$, and one gets $\alpha_k^{(n)}(t) = -i(h_k\dot g_k^{(n)*} - \dot h_k g_k^{(n)*})/2 $ and 
 $\beta_k^{(n)}(t)=-i (g_k^{(n)} \dot h_k - \dot g_k^{(n)} h_k)/2$.
  % \bea \label{systemcoefficients}\alpha_k^{(n)}(t) &=& \frac{1}{2i} (h_k\dot g_k^{(n)*} - \dot h_k g_k^{(n)*}) \nonumber \\
%\beta_k^{(n)}(t) &=& \frac{1}{2i} (g_k^{(n)} \dot h_k - \dot g_k^{(n)} h_k) \ . \eea
The operators $a^{\dagger}_{\vec{k}}(t)$ and  $ a_{\vec{k}}(t)$ are interpreted as annihilation and creation operators for real particles created in pairs from the vacuum $|0\rangle$  by the expanding universe. 

The number of created particles with momentum $\vec{k}$ is $\langle N_{\vec{k}}(t)\rangle \equiv \langle a^{\dagger}_{\vec{k}}(t)a_{\vec{k}}(t) \rangle = |\beta_k^{(n)}(t)|^2$, 
%and \be \label{Nsum}\langle N_{\vec{k}}(t)\rangle \equiv \langle a^{\dagger}_{\vec{k}}(t)a_{\vec{k}}(t) \rangle = |\beta_k^{(n)}(t)|^2 \ee
and the average number density of total created particles is
\be  \langle n (t)\rangle =\frac{1}{L^3 a^3} \sum_{\vec{k}} \langle N_{\vec{k}}(t)\rangle = \frac{1}{L^3 a^3} \sum_{\vec{k}} |\beta_k^{(n)} (t)|^2 \ . \label{numberpart}
 \ee
In order to have a well-defined expression for the mean number of created particles, we must use the minimum order $n$ that makes this quantity converge in the ultraviolet regime. Generically, the sum (\ref{numberpart}) is UV divergent for $n=0$, while it converges for $n=1$ (see also \cite{Fulling}). %We will see in the next section that, for the Bunch-Davies vacuum in de Sitter spacetime, we have in this regime $|\beta_k^{(0)}| \sim \mathcal{O} ( k^{-1})$ and $|\beta_k^{(1)}| \sim \mathcal{O} (k^{-2})$. 
Therefore, one needs in this case to use the Bogolubov coefficient  $\beta_k^{(1)} (t)$ } (this last criteria will change when considering spin-$1/2$ particles). Therefore, in the continuous limit $L \to \infty$, the number density of particles created at a given time $t$ is  
\be  \langle n (t)\rangle  = \frac{1}{2 \pi^2 a^3(t)} \int_{0}^{\infty} dk k^2    |\beta_k^{(1)} (t)|^2  \label{density} \ .\ee
% We note for completeness that, in order to have a finite uncertainty  for the particle number $(\Delta N_{\vec{k}})^2 \equiv \langle N_{\vec{k}}^2 \rangle - \langle N_{\vec{k}} \rangle^2 $, one would need the third adiabatic order.

The adiabatic expansion of the modes can be moved easily to an expansion of the 2-point function $\langle \phi (x) \phi (x') \rangle \equiv G(x, x')$ at coincidence $x=x'$. Using (\ref{phifield}), (\ref{phians}) and (\ref{ansatzWKB}), the adiabatic expansion of $G(x,x)$ in the continuous limit is written as
\bea \label{Adsub} G_{Ad}(x,x) & = &\frac{1}{2(2\pi)^3 a^3} \int d^3k [ w^{-1} + (W^{-1})^{(2)} \nonumber \\ & + & (W^{-1})^{(4)} + ... ] \ . \eea
$G(x,x)$ is formally a divergent quantity and must be renormalized. This is done in adiabatic renormalization by subtracting the expansion $ G_{Ad}(x,x)$ truncated to the minimal adiabatic order necessary to cancel all UV divergences that appear in the formal expression of the vacuum expectation value that one wants to compute. For instance, 
the computation of the renormalized variance $\langle \phi^2 \rangle $ requires  truncation up to second adiabatic order
\be \langle \phi^2 (x) \rangle_r = \frac{1}{2(2\pi)^3 a^3} \int d^3k [|h_k(t)|^2 - w^{-1} - (W^{-1})^{(2)} ] \ee
 while the renormalization of the stress-energy tensor needs subtraction up to fourth adiabatic order. Since it has been the observable more studied in the literature, we refer the reader interested in its full renormalization to the classical works \cite{parker-fulling74, bunch80}.

%%%%%%%%%%%%%%%%%%%%%%%%%%%%%%%%%%%%%%%%%%%%%%%%%%%%%%%%%%%%%%%%%%%%%%%%%%%
\section{Adiabatic expansion for spin one-half fields}

 With all the previous background on the adiabatic regularization method for scalars, we now enter into the main content of this work: its extension to spin-$1/2$ fields.
 
The covariant Dirac equation in curved spacetime is given by (see for instance \cite{parker-toms, birrell-davies})
\be i{\gamma}^{\mu}\nabla_{\mu} \psi - m\psi =0
\ee
where ${\gamma}^{\mu}(x) $ are the spacetime-dependent Dirac-matrices satisfying the condition $\{{\gamma}^{\mu}, {\gamma}^{\nu}\}= 2g^{\mu \nu}$ and  $\nabla_{\mu} \equiv \partial_{\mu} - \Gamma_{\mu}$ is the covariant derivative associated to the  spin connection $\Gamma_{\mu}$. 

Let us consider the spatially flat FLRW metric (\ref{frwmetric}). The matrices $\gamma^{\mu}(t)$ are related to  the constant Dirac matrices in Minkowski spacetime $\gamma^{\alpha}$, obeying  $\{{\gamma}^{\alpha}, {\gamma}^{\beta}\}= 2\eta^{\alpha \beta}$, by the simple relations
\be \gamma^0(t) = \gamma^0 \ ;  \  \ \ \ \ \ \ \   \gamma^i (t) = \gamma^i/a(t)\ee
Moreover, we also have $\gamma^{\mu}\Gamma_{\mu}= -3\dot a/2a \gamma_0$.
The Dirac equation is then of the form
\be (i\gamma^0\partial_0+ \frac{3i}{2}\frac{\dot a }{a}\gamma^0  +\frac{i}{a}\vec{\gamma}\vec{\nabla} -m)\psi=0 \ . \label{diracgamma} \ee
Let us now work with  the standard  Dirac-Pauli representation for the Dirac matrices 
\be 
\gamma^0 =
\left( {\begin{array}{cc}
 I & 0  \\
 0 & -I  \\
 \end{array} } \right)
\hspace{1cm} \vec\gamma = \left( {\begin{array}{cc}
 0 & \vec\sigma  \\
 -\vec\sigma & 0  \\
 \end{array} } \right) 
 \ee
 where $\vec{\sigma}$ are the usual Pauli matrices.
After  momentum expansion  
\be \psi=\sum_{\vec{k}} \psi_{\vec{k}}(t)e^{i\vec{k}\vec{x}} \ee it is convenient
%we have
%$(\partial_t+ 3\dot a/2a  +i\gamma^0\vec{\gamma}\vec{k}/a +i\gamma^0m)\psi_{\vec{k}}(t)=0$.
to write the Dirac field in terms of  two two-component spinors
\bea \label{psik}
&\psi_{\vec{k}}(t)=
\left( {\begin{array}{c}
  \frac{1}{\sqrt{L^3 a^3}}h^I_{{k}}(t) \xi_{\lambda} (\vec{k}) \\
  \frac{1}{\sqrt{L^3 a^3}}h^{II}_{{k}}(t)\frac{\vec{\sigma}\vec{k}}{k} \xi_{\lambda} (\vec{k})\\
 \end{array} } \right)
\eea
%The Dirac equation takes then the form 
%\bea (\partial_t+ \frac{3}{2}\frac{\dot a }{a} +im)f_{\vec{k}}^I=-i\frac{\vec{\sigma}\vec{k}}{a}f_{\vec{k}}^{II} \nonumber \\
%(\partial_t+ \frac{3}{2}\frac{\dot a }{a} -im)f_{\vec{k}}^{II}=-i\frac{\vec{\sigma}\vec{k}}{a}f_{\vec{k}}^I \ , \eea
where $ \xi_{\lambda} (\vec{k})$ is a constant  normalized two-component spinor $\xi_{\lambda}^{\dagger}\xi_{\lambda}=1$ such that  
 $\frac{\vec{\sigma}\vec{k}}{2k}\xi_{\lambda}= \lambda \xi_{\lambda}$.
$\lambda ={\pm} 1/2$  represents the eigenvalue for the helicity, or spin component along the $\vec{k}$ direction.
$h_{{k}}^I$ and $h_{{k}}^{II}$ are scalar functions,  which obey from (\ref{diracgamma}) the coupled first-order equations   
\bea \label{eqhI,II}&h_{{k}}^{II}=\frac{ia}{k}(\partial_t+im)h_{{k}}^I  \ \ , \ \  h_{{k}}^{I}=\frac{ia}{k}(\partial_t-im)h_{{k}}^{II} \label{eq:hII} , \eea
%&h_{{k}}^{I}=\frac{ia}{k}(\partial_t-im)h_{{k}}^{II}  \label{eq:hI}\ 
and the uncoupled second order equations:
\be \label{equationhI}(\partial_t^2 +\frac{\dot a}{a}\partial_t  +im\frac{\dot a}{a}+m^2 + \frac{k^2}{a^2} )h_{{k}}^I=0 \ , \ee 
and 
\be \label{equationhII}(\partial_t^2 +\frac{\dot a}{a}\partial_t  -im\frac{\dot a}{a}+m^2 + \frac{k^2}{a^2} )h_{{k}}^{II}=0 \ . \ee
%\be \label{equationhII}(\partial_t^2 +\frac{\dot a}{a}\partial_t  -im\frac{\dot a}{a}+m^2 + \frac{k^2}{a^2} )h_{{k}}^{II}=0 \ . \ee
The normalization condition for the four-spinor is 
\bea \label{normalization} &|h_{{k}}^I(t)|^2 +   |h_{{k}}^{II}(t)|^2 =1 \ . \eea

This condition guaranties the standard anticommutator relations for creation and annihilation operators  defined by the expansion
\be \psi= \sum_{\vec{k}} \sum_{\lambda = {\pm 1/2}} ( B_{\vec{k}, \lambda } u_{\vec{k}, \lambda}(t, \vec{x}) +  D^{\dagger}_{\vec{k}, \lambda} v_{\vec{k}, \lambda} (t, \vec{x}) )\ , \ee
where $u_{\vec{k}, \lambda}(t, \vec{x})$ is defined from an exact solution to the above equations 
\be u_{\vec{k}, \lambda} (\vec{x}, t) \equiv \frac{1}{\sqrt{L^{3}a^3}} e^{i\vec{k}\vec{x}}\left( {\begin{array}{c}
 h^I_{{k}}(t)  \xi_{\lambda}\\
h^{II}_{{k}}(t) \frac{\vec{\sigma}\vec{k}}{k}\xi_{\lambda}\\
 \end{array} } \right) \ . \ee 
These modes maintain the  standard normalization  with respect to the Dirac scalar product
\be (u_{\vec{k}, \lambda}, u_{\vec{k'}, \lambda'})= \int d^3x a^3{u^{\dagger}}_{\vec{k}, d\lambda}u_{\vec{k'}, \lambda'}= \delta (\vec{k}- \vec{k'}) \delta_{\lambda \lambda'}
\ee
 
The orthogonal modes $ v_{\vec{k}, \lambda} (t, \vec{x})$ are obtained by the charge conjugation operation $v_{\vec{k}, \lambda}=Cu_{\vec{k}, \lambda}=i\gamma^2u^*_{\vec{k}, \lambda}$. We then have
\bea \{B_{\vec{k}, \lambda}, B_{\vec{k}', \lambda'}^{\dagger}\}= \delta^3 (\vec{k}-\vec{k}')\delta_{\lambda \lambda'} \\  \{B_{\vec{k}, \lambda}, B_{\vec{k}', \lambda'}\}= 0=  \{B_{\vec{k}, \lambda}^{\dagger}, B_{\vec{k}', \lambda'}^{\dagger}\} \ , \eea
and similarly for the  $D_{\vec{k}, \lambda}$, $D_{\vec{k}, \lambda}^{\dagger}$ operators.

\subsection{ WKB-type expansions}

One could be tempted to  use the above  Klein-Gordon type Eqs. (\ref{equationhI}), (\ref{equationhII}) to generate a WKB-type expansion for $h_k^I$ and $h_k^{II}$. A redefinition of the field modes  as $\bar {h}^I_k \equiv a^{1/2}h_k^I$ and $\bar h^{II}_k \equiv a^{1/2}h_k^{II}$ converts those equations in
\be \label{equationbarhI}(\partial_t^2   +m^2 + \frac{k^2}{a^2} + \frac{\dot a^2}{4a^2} - \frac{\ddot a}{2a} +im\frac{\dot a}{a})\bar h_{{k}}^I=0 \ , \ee 
and 
\be \label{equationbarhII}(\partial_t^2   +m^2 + \frac{k^2}{a^2} + \frac{\dot a^2}{4a^2} - \frac{\ddot a}{2a}-im\frac{\dot a}{a})\bar h_{{k}}^{II}=0 \ . \ee
The WKB-type  ansatz (\ref{ansatzWKB}) works so well for scalar fields since it preserves  the Klein-Gordon product, and hence the Wronskian (\ref{Wronskian}). However, it  does not  preserve in general  the Dirac product and the associated Wronskian (normalization) condition (\ref{normalization}).

The presence of  a  complex quantity in the above Eqs. (\ref{equationbarhI}), (\ref{equationbarhII}) suggests  using  a generalized form of the WKB-type expansion \cite{Waterman, Kluger}
\be \label{WKBm}\bar h^{I, II}_k = \frac{N^{I, II}_k}{\sqrt {\Omega^{I, II}_k(t)}} e^{-i\int^t (\Omega^{I, II}_k(t') {\mp}i m \frac{\dot a(t')}{2a(t')\Omega^{I, II}_k(t')})dt'} \ , \ee
where $N^{I, II}_k$ are (time-independent) normalization constants to be fixed.  The Eqs. (\ref{equationbarhI}), (\ref{equationbarhII}) determine an adiabatic expansion of the form $\Omega_k^{I, II}= \omega (t) + \Omega^{I, II (1)}(t)+ \Omega^{I, II (2)}(t)+ ...$,  where  $\omega (t)=\sqrt{k^2/a^2 + m^2}$.   At  first adiabatic order one has $ \Omega^{I(1)}= 0= \Omega^{II(1)}$. The constants $N^{I, II}_k$ should be determined by imposing, order by order, the normalization condition $| h_{{k}}^I(t)|^2 +   | h_{{k}}^{II}(t)|^2 = 1$. It is easy to see that, at first adiabatic order, the constants $N^{I, II}_k$ cannot be fixed to fit this condition. %Note that in conformal time $\eta$ ($a d\eta = dt$) those equations have only second order derivatives with respect to $\eta$.
%\bea \label{normalization2} &&| h_{{k}}^I(t)|^2 +   | h_{{k}}^{II}(t)|^2 = a^{-1}(| \bar h_{{k}}^I(t)|^2 +   | \bar h_{{k}}^{II}(t)|^2 \nonumber \\
%&=& \frac{|N_k^I|^2}{a \omega (t)}e^{-2m\int^t \frac{\dot a(t')}{a(t')\omega(t')}dt'}+\frac{|N_k^{II}|^2}{a \omega (t)}e^{2m\int^t \frac{\dot a(t')}{a(t')\omega(t')}dt'} \nonumber  . \\
% \eea
 %We see immediately that the  constants $N^{I, II}_k$ cannot be fixed to fit the normalization $| h_{{k}}^I(t)|^2 +   | h_{{k}}^{II}(t)|^2 = 1$ up to second{\color{red}-order} adiabatic terms. 
 Only at  zeroth adiabatic order  we have a consistent  solution $N_k^I = \sqrt{a(\omega + m)/2}$, $N_k^{II}= \sqrt{a(\omega -m)/2}$.  %We find an obstruction to implement the expansion (\ref{WKBm}) at higher adiabatic orders. 
We should stress, nevertheless, that  the above ansatz is consistent for a spinor field in Minkowski space in the presence of a homogeneous time-dependent electric field \cite{Kluger}. To find an adiabatic expansion for spin-$1/2$ field modes in a FLRW universe we have to follow a different strategy.

We also remark that a WKB-type expansion can also be very useful to find approximate solutions to the Dirac equation in nontrivial backgrounds, as for instance in static, spherically symmetric spacetimes \cite{Groves}. However, the aim of those applications of the WKB-expansion are not directly linked to the proper renormalization expansion (in \cite{Groves} a DeWitt-Schwinger point-splitting expansion is used as the renormalization scheme). The strong requirements that an asymptotic expansion  needs to satisfy  to define a consistent renormalization scheme are not necessary for other purposes.

% If one carries out this strategy one gets unavoidable inconsistencies. 
%Therefore one should follow a different strategy.
\subsection{Adiabatic expansion for the spin-$1/2$ field modes}\label{adexpansionfermions}

In any case the zeroth adiabatic order should naturally generalize  the standard solution in Minkowski space
\bea h^{I M}_{{k}}(t) &=& \sqrt{\frac{\omega + m}{2\omega}} e^{-i\omega t} \nonumber \\
h^{II M}_{{k}}(t) &=& \sqrt{\frac{\omega - m}{2\omega}} e^{-i\omega t} \ , \eea
where here  $\omega=\sqrt{k^2 + m^2 }$. 
Therefore, the zeroth adiabatic  order must be of the form 
\bea g^{I (0)}_{{k}}(t) &=& \sqrt{\frac{\omega (t) + m}{2\omega(t)}} e^{-i\int^t \omega (t') dt'} \nonumber \\
g^{II(0)}_{{k}}(t) &=& \sqrt{\frac{\omega(t) - m}{2\omega(t)}} e^{-i\int^t\omega (t') dt'} \ , \eea
where from now on  $\omega (t) =\sqrt{k^2/a^2 + m^2 }$, as usual. It is easy to see that the zero order obeys the normalization condition $|g_{{k}}^{I (0)}(t)|^2 +   |g_{{k}}^{II(0)}(t)|^2 =1$.

The form of the above zeroth order modes and the structure of the field equations for $h_k^I$ and $h_k^{II}$  suggest the following 
template for the adiabatic expansion (truncated at adiabatic order $n$)
\bea   \label{nadiabatic}g^{I (n)}_{{k}}(t) &=& \sqrt{\frac{\omega  + m}{2\omega}} e^{-i\int^t (\omega (t') +\omega^{(1)}+ ... +\omega^{(n)})dt'}  \nonumber \\ & \times & (1+ F^{(1)}+... +F^{(n)}) \nonumber \\
 g^{II (n)}_{{k}}(t) &=& \sqrt{\frac{\omega - m}{2\omega}} e^{-i\int^t(\omega (t') + \omega^{(1)}+ ... +\omega^{(n)})dt'} \nonumber \\ & \times & (1+ G^{(1)}+... +G^{(n)}) \ , \eea
 where $\omega^{(n)}$, $F^{(n)}$ and $G^{(n)}$ are local functions of adiabatic order $n$.

\subsubsection{Adiabatic order $n=1$}

%\bea   \label{1adiabatic}g^{I (1)}_{{k}}(t) &=& \sqrt{\frac{\omega  + m}{2\omega}} e^{-i\int^t (\omega (t') +\omega^{(1)})dt'}   (1+ F^{(1)}) \nonumber \\
 %g^{II (1)}_{{k}}(t) &=& \sqrt{\frac{\omega - m}{2\omega}} e^{-i\int^t(\omega (t') + \omega^{(1)})dt'}  (1+ G^{(1)}) \ , \eea
At first adiabatic order, Eqs. (\ref{eqhI,II}) imply
\bea
&2a\omega^{2}\{(\mathfrak{Re}G^{\left(1\right)}-\mathfrak{Re}F^{\left(1\right)}+i(\mathfrak{Im}G^{\left(1\right)}-\mathfrak{Im}F^{\left(1\right)}))\left(m-\omega\right)\nonumber \\ &+\omega^{\left(1\right)}\}-im\left(m-\omega\right)\dot{a}  =  0\label{eq:sistad1} \ , \eea
and 
\bea 
&2a\omega^{2}\{(\mathfrak{Im}F^{\left(1\right)}+\mathfrak{Im}G^{\left(1\right)}+i(\mathfrak{Re}G^{\left(1\right)}-\mathfrak{Re}F^{\left(1\right)}))\left(m+\omega\right)\nonumber \\ &+i\omega^{\left(1\right)}\}+m\left(m+\omega\right)\dot{a}  =  0 \label{eq:sistad11} \ . 
\eea
Moreover, the modes at order $n=1$ should also respect the normalization condition at the given order
\be |g_{{k}}^{I (1)}(t)|^2 +   |g_{{k}}^{II(1)}(t)|^2 =1 \label{normalization1} \ , \ee
leading to the additional equation
\begin{equation}
1+\mathfrak{Re}G^{\left(1\right)}+\mathfrak{Re}F^{\left(1\right)}+\frac{m}{\omega}\left(\mathfrak{Re}F^{\left(1\right)}-\mathfrak{Re}G^{\left(1\right)}\right)=1\label{eq:normad1}
\ . \end{equation}
The solution to Eqs. (\ref{eq:sistad1}), (\ref{eq:sistad11}) and (\ref{normalization1}) is 
\begin{eqnarray}
\mathfrak{Re}F^{\left(1\right)} & = & 0 = \mathfrak{Re}G^{\left(1\right)} ,\label{eq:ref1} \nonumber \\
\omega^{\left(1\right)} & = & 0,\label{eq:w1} \nonumber \\
\mathfrak{Im}G^{\left(1\right)} & = & \mathfrak{Im}F^{\left(1\right)}+\frac{m}{2\omega^{2}}\frac{\dot{a}}{a}.\label{eq:imf1}
\end{eqnarray}

We see that  $\mathfrak{Im}F^{\left(1\right)}$ is undetermined. Nevertheless,   it is useful to realize  that  $F^{(1)}$ and $G^{(1)}$  can be parametrized as
\be F^{(1)}=-Ai \frac{m \dot a}{ \omega^2 a} + iK\frac{\dot a}{a} \,\,\,\,\,\,\,\,\,\,\,\, G^{(1)}= B i\frac{m \dot a}{ \omega^2 a}+ iK\frac{\dot a}{a} \label{efe1}\ee
where $A$ and $B$ are real arbitrary constants satisfying $A+B=1/2$, and $K(m, \omega)$ is an arbitrary functional of $m$ and $\omega$ with the adequate dimensions. 

\subsubsection{Adiabatic order $n=2$}

At second adiabatic order, Eqs. (\ref{eqhI,II}) restrict the form of the functions $F^{(2)}, G^{(2)}$ and $\omega^{(2)}$. We get
{\small \bea
&& \left(\left(\mathfrak{Re}F^{\left(2\right)}-\mathfrak{Re}G^{\left(2\right)}+i\left(\mathfrak{Im}F^{\left(2\right)}-\mathfrak{Im}G^{\left(2\right)}\right)\right)\left(m-\omega\right)  -\omega^{\left(2\right)}\right) \nonumber \\
&& \times 2a^{2}\omega^{4} +2\left(-2m^{2}+\omega^{2}\right){\left(\mathfrak{Im}F^{\left(1\right)}\right)'}\dot{a}^{2} \nonumber \\
&& +a\omega^{2}\left(\mathfrak{Im}F^{\left(1\right)}m\left(-m+\omega\right)\dot{a}+2{\left(\mathfrak{Im}F^{\left(1\right)}\right)'}\ddot{a}\right)=0 \ , \eea}
where $\left(\mathfrak{Im}F^{\left(1\right)}\right)'$ is a zeroth order adiabatic term defined by the expression 
\bea \dot{\left(\mathfrak{Im}F^{\left(1\right)}\right)}&=& \frac{d\mathfrak{Im}F^{\left(1\right)}}{d(\dot a/\omega^2 a)} \frac{d(\dot a/\omega^2 a)}{dt}\nonumber \\ &\equiv& \left(\mathfrak{Im}F^{\left(1\right)}\right)' \frac{d(\dot a/\omega^2 a)}{dt} \ , \eea
and
{\small \bea
&&
\left(\left(-\mathfrak{Re}F^{\left(2\right)}+\mathfrak{Re}G^{\left(2\right)}-i\left(\mathfrak{Im}F^{\left(2\right)}-\mathfrak{Im}G^{\left(2\right)}\right)\right)\left(m+\omega\right)+\omega^{\left(2\right)}\right)\nonumber \\
&& \times4a^{2}\omega^{4} +\dot{a}^{2}\left( m(m-2\omega)\omega+\right.\left.5m^{3}+4\left(2m^{2}-\omega^{2}\right){\left(\mathfrak{Im}F^{\left(1\right)}\right)'}\right) \nonumber \\
&& +2a\omega^{2}\left(\mathfrak{Im}F^{\left(1\right)}m\left(m+\omega\right)\dot{a}-\left(2{\left(\mathfrak{Im}F^{\left(1\right)}\right)'}\ddot{a}+m\right)\right) =0
\ . \eea}
Moreover, the normalization condition at second adiabatic order $\left|g_{k}^{I\:\left(2\right)}\left(t\right)\right|^{2}+\left|g_{k}^{II\:\left(2\right)}\left(t\right)\right|^{2}=1$
leads to
{\small \begin{eqnarray}
&&1+\mathfrak{Re}F^{\left(2\right)}+\mathfrak{Re}G^{\left(2\right)}+\frac{m}{\omega}\left(\mathfrak{Re}F^{\left(2\right)}-\mathfrak{Re}G^{\left(2\right)}\right) +\left(\mathfrak{Im}F^{\left(1\right)}\right)^{2}\nonumber \\
&&+\frac{m^{2}}{8\omega^{4}}\frac{\dot{a}^{2}}{a^{2}}\left(1-\frac{m}{\omega}\right)+\frac{m}{2\omega^{2}}\frac{\dot{a}}{a}\mathfrak{Im}F^{\left(1\right)}\left(1-\frac{m}{\omega}\right)  =  1\label{eq:sis3}
\end{eqnarray}
}

As for adiabatic order one, the solution to the above equations is not univocally fixed.  The general solution is given by
{\small 
\begin{eqnarray}
\mathfrak{Re}F^{\left(2\right)} & = & -\frac{1}{2}\left(\mathfrak{Im}F^{\left(1\right)}\right)^{2} \\
&+& \frac{m\left(m-\omega\right)\left(5m^{2}-2\omega^{2}\right)\dot{a}^{2}+2ma\omega\left(-m+\omega\right)\ddot{a}}{16a^{2}\omega^{6}}, \nonumber \\
\mathfrak{Re}G^{\left(2\right)} & = & -\frac{1}{2}\left(\mathfrak{Im}F^{\left(1\right)}\right)^{2} + \frac{m\left(-5m^{3}-5m^{2}\omega+2\omega^{3}\right)\dot{a}^{2}}{16a^{2}\omega^{6}} \nonumber \\
&+&\frac{2ma\omega^{2}\left(\left(m+\omega\right)\ddot{a}-4\omega^{2}\dot{a}\mathfrak{Im}F^{\left(1\right)}\right)}{16a^{2}\omega^{6}},\\
\omega^{\left(2\right)} & = & \frac{\dot{a}^{2}m\left(m-\omega\right)\left(5m^{2}+\left(m-2\omega\right)\omega\right)}{8a^{2}\omega^{5}}\\
\!\!\!\!\!\!\!\!\!\!\!\!\!\!\!\!\!\!\!\!\!\!\!\!\!\!\!\!\!\!\!\!\! &  & \!\!\!\!+\frac{8\dot a^2 \omega\left(-2m^{2}+\omega^{2}\right) {\left(\mathfrak{Im}F^{\left(1\right)}\right)'}}{8a^{2}\omega^{5}} \nonumber \\
&+&\frac{\ddot{a}\left(-m^{2}+\omega\left(m+4 {\left(\mathfrak{Im}F^{\left(1\right)}\right)'}\right)\right)}{8a^{2}\omega^{5}}\:,\nonumber \\
\!\!\!\!\!\!\!\!\!\!\!\!\!\!\!\!\!\!\!\!\!\!\!\!\!\!\!\!\!\!\mathfrak{Im}G^{\left(2\right)} & = & \mathfrak{Im}F^{\left(2\right)}.\label{eq:imf2} \end{eqnarray}}{\small \par} 
We note that, as for $\mathfrak{Im}F^{\left(1\right)}$,  $\mathfrak{Im}F^{\left(2\right)}$ is also undetermined.
%This ambiguity reflects the intrinsic ambiguity in the choice of the exact functions $h_k^I, h_k^{II}$ to characterize a given quantum state.

Before going to the third adiabatic order, it is convenient to analyze the expression for the adiabatic subtraction term involved in the renormalization of local observables requiring up to second adiabatic order. This is the case of $\langle \bar \psi \psi \rangle$. Given a vacuum state characterized by the exact mode functions $h_k^I, h_k^{II}$, the renormalized observable $\langle \bar \psi \psi \rangle_r$ is given by 
\bea &&\langle \bar \psi \psi \rangle_{r}= \frac{-2}{(2\pi)^3a^3} \times \nonumber \\
&& \int d^3k(|h_k^{I}|^2 - |h_k^{II}|^2- |g_k^{I (2)}|^2 + |g_k^{II (2)}|^2) \label{barpsipsi} \ .  \eea 
The explicit form of the subtraction terms in (\ref{barpsipsi}) is given by
\bea
 \label{subtractionbarpsipsi}|g_k^{I (2)}|^2 - |g_k^{II (2)}|^2 &=& \frac{m}{\omega}-m\frac{R}{24\omega^{3}} \nonumber \\
&+& \frac{m^{3}\left(7\dot{a}^{2}+2a\ddot{a}\right)}{8a^{3}\omega^{5}}-\frac{5m^{5}\dot{a}^{2}}{8a^{2}\omega^{7}} \ ,
 \eea
which turns out to be  independent of the ambiguity in $\mathfrak{Im}F^{\left(1\right)}$ and  $\mathfrak{Im}F^{\left(2\right)}$. This also happens for any  local observable, irrespective of the adiabatic order required in the renormalization (for further details see \cite{LMT}). Therefore, from now on we will fix the ambiguity by choosing  that $F^{(n)}(-m)= G^{(n)}(m)$ for every $n \geq 1$. This is equivalent to $g_k^{I (n)}(-m)= g_k^{II (n)}(m)$ and implies that $\mathfrak{Im}F^{\left(1\right)}= -\mathfrak{Im}G^{\left(1\right)}$ and $\mathfrak{Im}F^{\left(2\right)}= 0=\mathfrak{Im}G^{\left(2\right)}$, and hence 
{\small 
 \bea  \omega^{(1)}&=& 0  \\ 
  F^{(1)}&=& -i\frac{m\dot a}{4  \omega^2 a} \label{effe1} \\
     G^{(1)}&=& i\frac{m\dot a}{4  \omega^2 a} \ ,    \eea } 
 and   
\bea  \omega^{(2)}&=& \frac{5m^4\dot a^2 -3\omega^2m^2\dot a^2 -2\omega^2m^2\ddot a a}{8\omega^5a^2}  \\
  F^{(2)}&=& \frac{m^2R}{48\omega^4} -\frac{5m^4\dot a^2}{16\omega^6a^2}-\frac{m^2\dot a^2}{32\omega^4a^2} - \frac{mR}{48\omega^3} + \frac{5m^3\dot a^2}{16\omega^5 a^2} \label{effe2} \\
     G^{(2)}&=& \frac{m^2R}{48\omega^4} -\frac{5m^4\dot a^2}{16\omega^6a^2}-\frac{m^2\dot a^2}{32\omega^4a^2} + \frac{mR}{48\omega^3} - \frac{5m^3\dot a^2}{16\omega^5 a^2} \nonumber \ ,  \\   \eea 
where $R= 6(\ddot a/a + \dot a^2/a^2)$.

Let us finally remark that the first two terms in (\ref{subtractionbarpsipsi}) are, after integration in momenta, UV divergent. This is similar to what we have seen  in the renormalization of $\langle \phi^2 \rangle$. The first one is of zeroth adiabatic order and can be associated to the renormalization of the cosmological constant. The second one, of adiabatic order two, is proportional to the scalar curvature and it can be associated to the renormalization of  Newton's constant.  

\subsubsection{Third and fourth adiabatic order}

We can proceed in the same way to compute the solutions at third adiabatic order.  With the mentioned simplifying assumption $F^{(3)}(-m)= G^{(3)}(m)$ and after similar calculations, we get
{\small 
\begin{eqnarray}
\omega^{(3)}&=&0 \\
\mathfrak{Re}F^{\left(3\right)} & = & 0=\mathfrak{Re}G^{\left(3\right)} \\
\mathfrak{Im}F^{\left(3\right)} & = & - \mathfrak{Im}G^{\left(3\right)}=-\frac{m\left(-130m^{4}+97m^{2}\omega^{2}-8\omega^{4}\right)\dot{a}^{3}}{128a^{3}\omega^{8}}\:\label{eq:singad3} \nonumber \\
&-&\frac{(4ma\omega^{2}(19m^{2}-8\omega^{2})\dot{a}\ddot{a}-8ma^{2}\omega^{4}\dddot{a}}{128a^{3}\omega^{8}} \nonumber \\
\end{eqnarray}
}{\small \par}

Finally, the fourth-order contributions are given in appendix A. We can continue the iteration indefinitely for all adiabatic orders, but relevant observables require at most subtractions up to fourth adiabatic order.
 
\section{Adiabatic regularization for spin one-half fields}

 Having developed the extended adiabatic expansion for spin-$1/2$ fields, we move now to its application for the obtention of conformal anomalies and the number operator.

\subsection{Anomalies for spin one-half fields}

The purpose of this section is to prove the consistency of the proposed adiabatic expansion for spin-$1/2$ field modes by working out the conformal and axial anomalies  in a FLRW spacetime. We extend in this way the adiabatic regularization to spin-$1/2$ fields.  We will find exact agreement with those obtained from other renormalization methods. 

\subsubsection{Conformal anomaly}

The stress-energy tensor of the Dirac field in a curved background can be expressed, using the Dirac equation  $i{\gamma}^{\mu}\nabla_{\mu} \psi - m\psi =0$, as  
\bea T_{\mu\nu}&=& \frac{i}{2} \left[\bar \psi \gamma_{(\mu}\nabla_{\nu)}\psi - (\nabla_{(\mu}\bar \psi)\gamma_{\nu)} \psi \right] 
%\nonumber \\
%&-&g_{\mu\nu}\left[ \frac{i}{2} (\bar\psi\gamma^{\rho}\nabla_{\rho}\psi - \nabla_{\rho}\bar\psi \gamma^{\rho}\psi ) - m\bar \psi \psi \right] 
\ . \eea
One obtains immediately that the trace of the stress-energy tensor takes the simple form
\be T^{\mu}_{\mu}= m\bar \psi \psi \ . \label{trace} \ee

When the field is massless the trace vanishes, signaling the emergence of the conformal invariance. However, in the quantum theory the expectation value  $\langle T^{\mu}_{\mu} \rangle = m \langle \bar \psi \psi \rangle$ takes a nonzero value even in the massless limit. Our purpose is to perform the calculation of this anomalous trace using the extension of the adiabatic regularization method for spin-$1/2$ introduced above. Since the expectation value $ \langle \bar \psi \psi \rangle$ is now regarded as a piece of the average value of the stress-energy tensor $\langle T_{\mu\nu} \rangle$, the renormalization should be performed up to the fourth adiabatic order. Taking into account that 
\bea &&\langle \bar \psi \psi \rangle_{r}= \frac{-2}{(2\pi)^3a^3} \times \nonumber \\
&& \int d^3k(|h_k^{I}|^2 - |h_k^{II}|^2- |g_k^{I (4)}|^2 + |g_k^{II (4)}|^2) \label{tracer} \ ,  \eea 
one can evaluate the trace anomaly by taking the massless limit in the above expression
\bea  \langle T_{\mu}^{\mu} \rangle_r &=& \lim_{m \to 0} \frac{-2 m}{(2 \pi)^3 a^3} \int d^3 k (|h_k^I|^2 - |h_k^{II}|^2  \nonumber \\
 &-& |g_k^{I(4)}|^2 + |g_k^{II(4)}|^2)  \ . \nonumber \eea
Only the fourth-order adiabatic subtraction terms survive in limit $m \to 0$. Therefore, the trace anomaly is given by the massless limit of the following integral 
{\small \begin{eqnarray}
 & & \frac{2m}{\left(2\pi a\right)^{3}}\int d^{3}k\left(\frac{m\dot{a}^{4}+11ma\dot{a}^{2}\ddot{a}+4ma^{2}\ddot{a}^{2}+7ma^{2}\dot{a}\dddot{a}+ma^{3}\ddddot{a}}{32a^{4}\omega^{5}}\right. \nonumber \\
\!\!\!\!\!\!\!\!\!\!\!\!\!\!\!\!\!\!\!\! &  & \ \ \ \ \!\!\!\!\!\!+\frac{-86m^{3}\dot{a}^{4}-211m^{3}\dot{a}^{2}\ddot{a}-29m^{3}a^{2}\ddot{a}^{2}-42m^{3}a^{2}\dot{a}\dddot{a}+ma^{3}\ddddot{a}}{32a^{4}\omega^{7}}\nonumber \\
\!\!\!\!\!\!\!\!\!\!\!\!\!\!\!\!\!\!\!\!&  & \ \ \ \ \!\!\!\!\!\!+\frac{6636m^{5}a^{2}\dot{a}^{4}+6720m^{5}a^{3}\dot{a}^{2}\ddot{a}+336m^{5}a^{4}\ddot{a}^{2}+448m^{5}a^{4}\dot{a}\dddot{a}}{32a^{6}\omega^{9}}\nonumber \\
\!\!\!\!\!\!\!\!\!\!\!\!\!\!\!\!\!\!\!\! &  & \ \ \ \ \ \!\!\!\!\!\!\!\!\left.+\frac{-158592m^{7}a^{4}\dot{a}^{4}-59136m^{7}a^{5}\dot{a}^{2}\ddot{a}}{8192a^{8}\omega^{11}}+\frac{1155m^{9}\dot{a}^{4}}{128a^{4}\omega^{13}}\right)\: . 
\end{eqnarray}
}
This integral is finite by construction and can be worked out analytically. The trace anomaly is then 
{\small
\bea
\left\langle T_{\mu}^{\:\mu}\right\rangle _{r}&=&\lim_{m\rightarrow0}\frac{4m}{\left(2\pi\right)^{2}a^{3}}\left[\frac{-4\dot{a}^{2}\ddot{a}+9a\dot{a}\dddot{a}+3a\left(\ddot{a}^{2}+a\ddddot{a}\right)}{240m}\right]\nonumber \\
&=& \frac{-4\dot{a}^{2}\ddot{a}+9a\dot{a}\dddot{a}+3a\left(\ddot{a}^{2}+a\ddddot{a}\right)}{240 \pi^2 } \label{eq:intertrace}
\eea}
This result can be rewritten as a linear combination of the covariant scalars 
\begin{eqnarray}
R^{2} & = & \left[6\left(\frac{\dot{a}^{2}}{a^2}+\frac{\ddot{a}}{a}\right)\right]^{2}\:,\\
\square R & = & 6\left(\frac{\ddot{a}}{a}+\frac{\ddddot{a}}{a}-\frac{5\dot{a}^{2}\ddot{a}}{a^{3}}+\frac{3\dot{a}\dddot{a}}{a^{2}}\right)\:,\\
R_{\mu\nu}R^{\mu\nu} & = & 12\left(\frac{\dot{a}^{4}}{a^{4}}+\frac{\ddot{a}^{2}}{a^{2}}+\frac{\dot{a}^{2}\ddot{a}}{a^{3}}\right)\:.
\end{eqnarray}
We find
\begin{eqnarray}
\left\langle T_{\mu}^{\:\mu}\right\rangle _{r} & = & \frac{1}{2880\pi^{2}}\left[-11\left(R_{\alpha\beta}R^{\alpha\beta}-\frac{1}{3}R^{2}\right)+6\,\square R\right] \nonumber \\
 & = & \frac{1}{2880\pi^{2}}\left[\frac{11}{2}G+6\,\square R\right] \label{eq:anomtracead}\ , 
\end{eqnarray}
where in the second line we have introduced the Gauss-Bonnet invariant $G$, which for a FLRW spacetime is given by $G= -2(R_{\mu\nu}R^{\mu\nu}-R^2/3)$. 
The conformal anomaly is generically given for a conformal field of  spin $0, 1/2$ or $1$ in terms of three parameters
\bea
&& \left\langle T_{\mu}^{\:\mu}\right\rangle _{r}=\frac{1}{2880\pi^{2}}\left(AC_{\mu\nu\rho\sigma}C^{\mu\nu\rho\sigma}+BG +C\square R \right)\ . \,\,\, \label{eq:anomtrace}
\eea

The result obtained for a Dirac spin-$1/2$ field by other renormalization procedures is $A=-9, B=11/2, C=6$ \cite{birrell-davies}. Our above result (\ref{eq:anomtracead}) agrees exactly with the results obtained from other methods.  We note that in a FLRW spacetime the conformal tensor $C_{\mu\nu\rho\sigma}$ vanishes identically. 
%This can be regarded as a nontrivial test of the robustness of our proposal.

We stress that no $R^2$ term appears in (\ref{eq:anomtracead}) and (\ref{eq:anomtrace}), although such a term  could have appeared. The vanishing of an $R^2$ term when the  trace
anomaly is expressed in terms of $G$, $\Box R$ and  $C_{\mu\nu\rho\sigma}C^{\mu\nu\rho\sigma}$ is required by consistency with the theorem that no creation of particles obeying conformally invariant equations occurs in an FLRW expanding universe \cite{parker66, parker69}. This is the 
case for a massless spin-$1/2$ field. This theorem is based on the conformal invariance of the field equations and it is respected by the conformal anomaly, as shown in \cite{Parker79}.. [For physical implications of this fact for the electromagnetic field see \cite{agullo-navarro13}].

\subsubsection{Axial anomaly}

In curved spacetime the axial vector current $J^{\mu}_A\equiv \bar \psi \gamma^{\mu}\gamma^5 \psi$, where $\gamma^5 \equiv i\gamma^0\gamma^1\gamma^2\gamma^3$,  obeys the covariant equation $\nabla_{\mu} J^{\mu}_A= 2im\bar \psi \gamma^5 \psi$. For a massless Dirac field  the classical axial current is conserved, due to the chiral symmetry. At the quantum level the expectation value
$\langle \nabla_{\mu} J^{\mu}_A \rangle$ may acquire a nonzero value in the massless limit. We want to evaluate this quantity using the adiabatic regularization for fermions. The strategy is similar to the evaluation of the conformal or trace anomaly. Since the divergences of  $\langle \nabla_{\mu} J^{\mu}_A \rangle$ are of fourth adiabatic order we have to work out  $\langle 2im\bar \psi \gamma^5 \psi \rangle $ also at fourth adiabatic order. In this case
  
\bea \langle \bar{\psi} \gamma^5 \psi \rangle^{(4)}_r &=& \frac{-2}{(2 \pi)^3 a^3}  \int d^3 k (h_k^{I*} h_k^{II} - h_k^{II*} h_k^{I}  \nonumber \\ &-& g_k^{I (4)*} g_k^{II (4)}+ g_k^{II (4)*} g_k^{I (4)} ) \ . \eea
Keeping only those terms that survive in the massless limit  we have
{\small \bea
&&\left\langle \overline{\psi}\gamma{5}\psi\right\rangle _{r} = \frac{-2}{\left(2\pi\right)^{3}a^{3}}\int d^{3}k\frac{imk}{16a^{3}\omega^{9}}\biggl(\dot{a}^{3}\left(-35m^{4} \right.\nonumber \\
&+&\left. 25m^{2}\omega^{2}-2\omega^{4}\right) +4a\omega^{2}\dot{a}\ddot{a}\left(5m^{2}-2\omega^{2}\right)+2a^{2}\omega^{4}\dddot{a}\biggr) + O(m) \nonumber  
\ . \eea
}
The integral is finite and can be computed analytically. We find
\bea \label{axial}\langle \nabla_{\mu} J^{\mu}_A \rangle_r &=& \lim_{m \to 0} 2im \langle \bar \psi \gamma^5 \psi \rangle \nonumber \\
&=& \lim_{m \to 0} 2im \frac{-ia(2\dot a\ddot a + a \dddot a)}{12} =0 \ . \eea
The vanishing of the axial current anomaly in our FLRW spacetime agrees with the result obtained from other renormalization methods. In a general spacetime the axial anomaly is given by (see, for instance, \cite{parker-toms})
\be \label{axiala}\langle \nabla_{\mu} J^{\mu}_A \rangle_r = \frac{1}{384\pi^2} \epsilon^{\mu\nu\rho\sigma}R_{\mu\nu}^{\ \  \lambda \xi}R_{\rho\sigma\lambda\xi} \ . \ee
It is very easy to check that for a FLRW spacetime the right-hand side of (\ref{axiala}) vanishes identically, in agreement with our result (\ref{axial}).

\subsection{Number operator} 

Let us analyze  the number operator for spin-$1/2$ Dirac particles. As for bosons, the quantized field $\psi$ can also be expanded in terms of the fermionic adiabatic modes $g_{\vec{k}, \lambda}(\vec{x}, t)$, 
\be \psi= \sum_{\vec{k}, \lambda}  ( b_{\vec{k}, \lambda }(t) g^{(n)}_{\vec{k}, \lambda}(t, \vec{x}) +  d^{\dagger}_{\vec{k}, \lambda}(t) g^{c (n)}_{\vec{k}, \lambda} (t, \vec{x}) ) \ee
  where  $g^c_{\vec{k}, \lambda} (t, \vec{x}) $ are the corresponding adiabatic modes obtained by the charge conjugation operation $C$ and
  \be   g^{(n)}_{\vec{k}, \lambda}(t, \vec{x}) =  \frac{1}{\sqrt{L^{3}a^3}} e^{i\vec{k}\vec{x}}\left( {\begin{array}{c}
  g^{I (n)}_{{k}}(t) \xi_{\lambda}\\
 g^{II (n)}_{{k}}(t)\frac{\vec{\sigma}\vec{k}}{k}\xi_{\lambda}\\
 \end{array} } \right) \ .\ee 
The Bogolubov coefficients can be now obtained from the exact modes $h_k^I(t)$ and $h_k^{II}(t)$ by solving the following system of equations:
\bea \label{systemhdothfermions}h^I_k(t) &=& \alpha^{(n)}_k(t) g_{{k}}^{I(n)}(t) - \beta^{(n)}_k(t) g_{{k}}^{II (n)}(t) \nonumber \\
  h^{II}_k(t) &=& \alpha^{(n)}_k(t)  g_{{k}}^{II (n)}(t) + \beta^{(n)}_k(t) g_{{k}}^{I(n)}(t)  \ . \eea 
 We have restricted for simplicity to the $\lambda=1/2$ case; similar equations apply for the opposite helicity. The solution of this system is, using that $g_k^{I (n)}$ and $g_k^{II (n)}$ follow the normalization condition (\ref{normalization}),
\bea
\beta_k^{(n)} (t) & = & g_k^{I (n)} h_k^{II}- g_k^{II (n)}  h_k^{I}  \nonumber \\
\alpha_k^{(n)} (t) & = & g_k^{I(n)*} h_k^{I} + g_k^{II (n)*} h_k^{II}  \label{0fercoef} \ .
\eea 
These Bogolubov coefficients obey the relation $|\alpha^{(n)}_k(t)|^2+ |\beta^{(n)}_k(t)|^2=1$ up to order $n$. On the other hand, the average number of created fermionic particles of specific helicity and charge with momentum $\vec{k}$ is
\be \langle N_{\vec{k}} \rangle= \langle b^{\dagger}_{\vec{k}}(t)b_{\vec{k}}(t) \rangle = |\beta^{(n)}_k(t)|^2 \ . \ee

As for bosons, we must use the minimum order that makes this integral converge in the ultraviolet limit. It is generally found that for large $k$, $|\beta_k^{(0)}(t)| \sim  \mathcal{O} (k^{-2})$. This is confirmed in the next section for de Sitter spacetime. This behavior guarantees the finiteness of the average number density of created particles when summed for all momenta:
\be \frac{1}{L^3 a^3} \sum_{\vec{k}} \langle N_{\vec{k}}(t)\rangle = \frac{1}{L^3 a^3} \sum_{\vec{k}} |\beta_k^{(0)}|^2 < \infty \label{numberpart3} \ee
We note that, in contrast with the scalar field, this result is obtained with the zeroth adiabatic order. [We note for completeness that in the calculation of the uncertainty for the particle number,  one would need $\beta_k^{(2)} (t)$ to have an UV finite result].

Finally, returning to the continuous limit, the density of spin one-half particles as a function of time of specific charge and helicity is
\be  \langle n (t)\rangle  = \frac{1}{2 \pi^2 a^3(t)} \int_{0}^{\infty} dk k^2    |\beta_k^{(0)} (t)|^2  \label{densityfer} \ .\ee

%The uncertainty is $\Delta  N_{\vec{k}} = \langle b^{\dagger}_{\vec{k}}(t)b_{\vec{k}}(t) b^{\dagger}_{\vec{k}}(t) b_{\vec{k}}(t) \rangle$. We need to use the second order coefficients when obtaining the total uncertainty
%\be \Delta  N_{\vec{k}} = \langle b^{\dagger}_{\vec{k}}(t)b_{\vec{k}}(t) b^{\dagger}_{\vec{k}}(t) b_{\vec{k}}(t) \rangle = \sqrt{ |\beta^{(2)}_k(t)|^2 - |\beta^{(2)}_k(t)|^4} \ee .
%because in de Sitter $|\beta_k^{(1)}| \sim O(k^{-3})$ and  $|\beta_k^{(2)}| \sim O(k^{-4})$. Therefore, we must use the second order coefficient to ensure a convergent uncertainty in the ultraviolet regime.
%\be \frac{1}{L^3 a^3} \sum_{\vec{k}} \Delta  N_{\vec{k}} = \frac{1}{L^3 a^3} \sum_{\vec{k}} \sqrt{ |\beta^{(2)}_k(t)|^2 - |\beta^{(2)}_k(t)|^4} < \infty \label{numberpart2} \ee

\section{Spin one-half field in de Sitter spacetime}

We analyze in this section the particle creation and the renormalized stress-energy tensor for a spin-$1/2$ field in de Sitter spacetime. This space is described by the metric (\ref{frwmetric}) with scale factor
\be a(t) = e^{H t} \label{scalesitter} \ee
and $H$ constant. The coupled differential Eqs. (\ref{eqhI,II}) take the form
 \bea & h_{{k}}^{II}= \frac{i e^{H t}}{k} ( \partial_t + i m ) h_{k}^I  \ , \  h_{k}^{I}= \frac{i e^{H t}}{k}( \partial_t - i m )h_{k}^{II} , \label{HK3} \eea
while the uncoupled Eqs. (\ref{equationhI}) and (\ref{equationhII}) are
\be \left( \frac{\partial^2 }{\partial t^2} + H \frac{\partial }{\partial t} + i m H + \frac{k^2}{e^{2Ht}} + m^2 \right) h_k^I = 0 \ , \label{HK1} \ee
\be \left( \frac{\partial^2 }{\partial t^2} + H \frac{\partial }{\partial t} - i m H + \frac{k^2}{e^{2Ht}} + m^2 \right) h_k^{II} = 0 \label{HK2} \ .\ee
It is helpful to define the following dimensionless variables as
\be z \equiv k H^{-1} e^{-H t} \,\,\,\,\,\,\,\,\,\,\,\, \mu \equiv \frac{m}{H} \ .\ee
In terms of these variables, the solution of (\ref{HK1}) is given by the two independent functions $\sqrt{z} H_{\frac{1}{2} - i \mu}^{(1)} (z)$ and $\sqrt{z} H_{\frac{1}{2} - i \mu}^{(2)} (z)$, but only the first one satisfies the condition  $h_k^{I}  \overset{{\scriptstyle t\rightarrow -\infty}}{\sim} g_k^{I (n)} $  (where the order $n$ is arbitrary). In the same way, Eq. (\ref{HK2}) has two independent solutions  $\sqrt{z} H_{- \frac{1}{2} - i \mu}^{(1)} (z)$ and $\sqrt{z} H_{- \frac{1}{2} - i \mu}^{(2)} (z)$ but only the first one obeys $h_k^{II}  \overset{{\scriptstyle t\rightarrow -\infty}}{\sim} g_k^{II (n)} $. Therefore, we take
 \be h_k^{I}= i \alpha \frac{\sqrt{ \pi z}}{2} e^{\frac{\pi \mu}{2}} H^{(1)}_{\frac{1}{2} - i \mu}(z) \label{hdesitter1} \ ,\ee 
and 
\be h_k^{II} = \beta \frac{\sqrt{ \pi z}}{2} e^{\frac{\pi \mu}{2}} H^{(1)}_{- \frac{1}{2} - i \mu}(z) \label{hdesitter2} \ee 
where $\alpha$ and $\beta$ are real constants to be fixed and the overall normalization factor $\sqrt{\pi z}/2$ has been extracted for convenience. By substituting (\ref{hdesitter1}) and (\ref{hdesitter2}) into (\ref{HK3}), we find $\alpha=\beta$. Finally, by imposing the normalization condition (\ref{normalization}), we find $|\alpha| = 1$. Therefore, we have
\be h_k^{I}= i\frac{\sqrt{ \pi z}}{2} e^{\frac{\pi \mu}{2}} H^{(1)}_{\frac{1}{2} - i \mu}(z) \label{hdesitter3} \ee
and 
\be h_k^{II} = \frac{\sqrt{ \pi z}}{2} e^{\frac{\pi \mu}{2}} H^{(1)}_{- \frac{1}{2} - i \mu}(z) \label{hdesitter4} \ee 
 up to a constant phase factor. Equations (\ref{hdesitter3}) and (\ref{hdesitter4}) determine  a vacuum for spin one-half fields analogous to the Bunch-Davies vacuum \cite{Bunch-Davies} for scalars, because it is the solution that reproduces the adiabatic modes for initial times. 

We note that $h_k^{II} (m) = h_k^I (-m)$. This is seen more clearly if we write (\ref{hdesitter3}) and (\ref{hdesitter4}) using the property $H_{- \nu}^{(1)} (z) = (-1)^{\nu} H_{\nu}^{(1)} (z)$ as
\bea
h_k^{I} &=&  \frac{\sqrt{\pi z}}{4} \left( i  e^{\frac{\pi \mu}{2}} H_{\frac{1}{2} - i \mu}^{(1)} (z)  + e^{- \frac{\pi \mu}{2}}  H_{-\frac{1}{2} + i \mu}^{(1)} (z) \right) \nonumber \\
h_k^{II} &=& \frac{\sqrt{\pi z}}{4} \left(  e^{\frac{\pi \mu}{2}} H_{-\frac{1}{2} - i \mu}^{(1)} (z) + i e^{ - \frac{\pi \mu}{2}} H_{\frac{1}{2} + i \mu}^{(1)} (z)  \right) \ .\nonumber \\
& &  
\eea
%A similar symmetry property $g_k^{I}(-m)= g_k^{II}(m)$ has been used in the adiabatic expansion of modes (Section \ref{adexpansionfermions})  to fix it univocally.
 
The full orthonormalized spinors $u_{\vec{k}, \lambda} (\vec{x}, t)$ and  $v_{\vec{k}, \lambda} (\vec{x}, t)$ for de Sitter space can be constructed as
\be u_{\vec{k}, \lambda} (\vec{x}, t) \equiv \frac{i \sqrt{z}}{4 \pi a^{3/2}} e^{i\vec{k}\vec{x}} e^{\frac{\pi \mu}{2}} \left( {\begin{array}{c} 
H^{(1)}_{\nu}(z)  \xi_{\lambda}\\
-iH^{(1)}_{\nu-1}(z) \frac{\vec{\sigma}\vec{k}}{k}\xi_{\lambda}\\
\end{array} } \right) \ee 
   \be v_{\vec{k}, \lambda} (\vec{x}, t) \equiv  \frac{\sqrt{z}}{4 \pi a^{3/2}} e^{i\vec{k}\vec{x}} e^{\frac{\pi \mu}{2}} \left( {\begin{array}{c}
-i H^{(1)*}_{\nu-1}(z) \frac{\vec{\sigma}\vec{k}}{k}\xi_{\lambda}\\
H^{(1)*}_{\nu}(z) \xi_{\lambda}\\
\end{array} } \right) \ee 
where we have defined the $\nu$ coefficient as
 \be \nu \equiv \frac{1}{2} - i \mu \ .\ee 

In order to study the particle number operator and the stress-energy tensor, we need the first terms  of the adiabatic expansion  introduced in Sec. III and particularized for (\ref{scalesitter}). We assume for simplicity the condition $F^{(n)} (m) = G^{(n)} (-m)$. If we substitute $\omega = H (z^2 + \mu^2)^{1/2}$, $\dot{a}/a = H$ and $R=12 H^2$ into (\ref{effe1}) and (\ref{effe2}), we obtain
\be F^{(1)} = \frac{-i \mu}{4 (z^2 + \mu^2)} \label{desitterf1}\ee and
\bea
F^{(2)} = &-& \frac{\mu}{4 (z^2 + \mu^2)^{3/2}} + \frac{7 \mu^2}{32 (z^2 + \mu^2)^2} \\ &+& \frac{5 \mu^3}{16 (z^2 + \mu^2)^{5/2}} -  \frac{5 \mu^4}{16 (z^2 + \mu^2 )^3} \ . \label{desitterf2}
\eea
The third- and fourth-order contributions can be obtained in a similar way, but they are not explicitly written here. 

We move now to the analysis of the particle number and the stress-energy tensor in de Sitter spacetime.

\subsection{Particle creation}

 The $\beta_k^{(n)} (t)$ Bogolubov coefficient (\ref{0fercoef}) for de Sitter space is found to be:
\bea
&& \beta_k^{(n)} (t) = e^{-i \int^t \omega (t') + \dots \omega^{(n)} (t')} \frac{\sqrt{\pi z}}{2^{3/2} (z^2 + \mu^2)^{1/4}} e^{\frac{\pi \mu}{2}} \times \nonumber\\
&& \left\{ (1 + \dots + F^{(n)}) H_{\nu - 1}^{(1)} (z) \sqrt{(z^2 + \mu^2)^{1/2} + \mu} \right. - \nonumber \\
&& \left. (1 + \dots + G^{(n)}) i H_{\nu}^{(1)} (z) \sqrt{(z^2 + \mu^2)^{1/2} - \mu} \right\} \ .
\eea 
 We find that in the ultraviolet limit
\be |\beta_k^{(n)}| \overset{{\scriptstyle k\rightarrow \infty}}{\sim} \mathcal{O} \left( \frac{1}{k^{n+2}} \right) \ . \label{asympfer}\ee
 Therefore, for spin-$1/2$ fields the zeroth adiabatic order suffices to give a UV finite expectation value for the number density and is given by Eq. (\ref{densityfer}), which is written in terms of $z$ as
\be \langle n \rangle_f = \frac{H^3}{2 \pi^2} \int_0^{\infty} dz z^2 |\beta_k^{(0)}|^2 \label{density-desitter} \ . \ee
This integral can be rewritten more conveniently as
\be \langle n \rangle_f = \frac{H^3}{16 \pi} e^{\pi \mu} J (\mu) \label{density-desitter2} \ , \ee
where $J (\mu)$ is a function given by the following expression:
\bea
J(\mu) &\equiv & \int_0^{\infty} dz \frac{z^3}{\sqrt{z^2 + \mu^2}} \times \left|  H_{\nu - 1}^{(1)} (z) \sqrt{(z^2 + \mu^2)^{1/2} + \mu} \right. \nonumber \\ 
&-& \left. i H_{\nu}^{(1)} (z) \sqrt{(z^2 + \mu^2)^{1/2} - \mu}\right|^2 \  . \label{jmu}  
\eea 

 It is useful to compare the fermionic density (\ref{density-desitter2}) with the scalar one, which is obtained with the methods sketched in Sec. II. The solution of (\ref{hqeq}) with scale factor (\ref{scalesitter}) that corresponds to the usual Bunch-Davies vacuum is $ h_k = \sqrt{(\pi /2 H )} e^{-\pi \mathfrak{Im} \nu /2} H_{\nu}^{(1)} (z)$ with $\nu \equiv \sqrt{(9/4) - \mu^2}$ (we take $\xi = 0$). The Bogolubov coefficients associated to this state are found to behave as $ |\beta_k^{(n)}| \overset{{\scriptstyle k\rightarrow \infty}}{\sim} \mathcal{O} ({k^{-n-1}})$, so Eq. (\ref{density}) holds. After some algebra, we find that the bosonic density can be written as
\be \langle n\rangle_b = \frac{H^3 e^{- \pi \mathfrak{Im} \nu}}{16 \pi} I (\mu) \label{density-bosons} \ee 
where $I (\mu)$ is a function given by the integral
\be
\begin{split}
I (\mu) & \equiv \int_0^{\infty} dv \frac{v^2}{\sqrt{v^2 + \mu^2}} \left| v H_{\nu - 1}^{(1)} (v) - \right. \\ & \left. \left( \nu + i \sqrt{v^2 + \mu^2} - \frac{v^2}{2 (v^2 + \mu^2)} \right) H_{\nu}^{(1)} (v) \right|^2 \ .\label{imu}
\end{split}
\ee
By construction, the integrals $J (\mu)$ and $I(\mu)$ are convergent in the ultraviolet regime. Also, densities (\ref{density-desitter2}) and (\ref{density-bosons}) are time-independent because de Sitter is a maximally symmetric spacetime with no preferred coordinate points.

\begin{figure}
\begin{center}
\includegraphics[width=7.7cm]{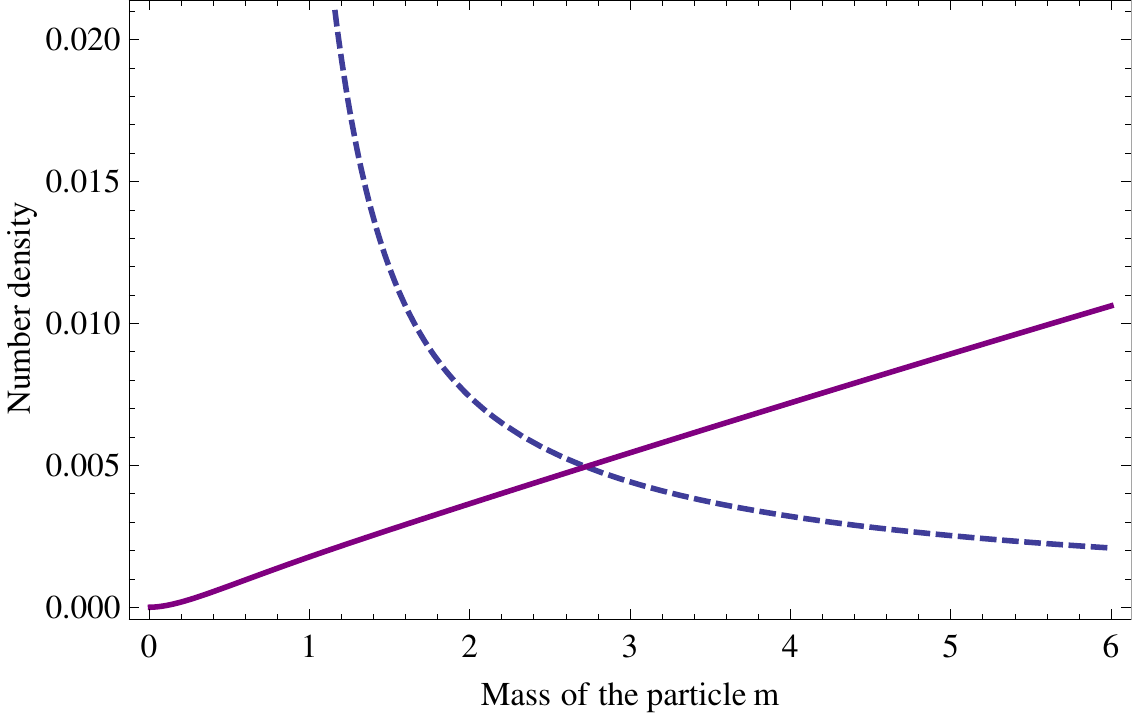} 
\end{center}
\caption{Number densities for fermions $\langle n \rangle_f$  (continuous line) and bosons $\langle n \rangle_b$  (dashed line) as a function of the mass $m = \mu H$ for $H=1$.}
\label{figura2}
\end{figure}

Figure \ref{figura2} shows the average densities (\ref{density-desitter2}) and (\ref{density-bosons}) as a function of the particle mass for $H=1$. Their behaviors are quite different. For bosons, one finds that $\lim_{\mu \to \infty} \langle n \rangle_b = 0$, and that $\langle n \rangle_b$ has an infrared divergence in the massless case $\lim_{\mu \to 0} \langle n \rangle_b = \infty$. [However, this divergence is somewhat spurious, because it appears for massless particles with no momentum, which do not contribute to the total energy content. This is confirmed because the bosonic $\langle T_{\mu \nu} \rangle$ obtained in \cite{Dowker-Critchley76} and \cite{Bunch-Davies} is finite in this limit].

Nevertheless, the behavior of $\langle n \rangle_f$ is quite different. First, we find that particle creation does not happen for massless fermions. The statement that creation of massless fermions is forbidden is quite general. It is due to the conformal invariance of the field theory, because creation of particles does not happen in any conformally invariant theory in a conformally flat metric such as (\ref{frwmetric}) .  As already stressed, this is compatible with the conformal anomaly. 

 On the other hand, it is found that $\langle n \rangle$ grows linearly with the mass. More specifically, for $m \gg H$, $\langle n \rangle$ grows as $\langle n \rangle \sim \alpha m H^2$  with $\alpha$ a constant. Therefore, one would expect that for large masses the effects of spontaneous particle creation can be so important that backreaction effects must be taken into account. The final result could be an instability of de Sitter space. However, our assumed quantum state conspires to protect this to happen, as we will shortly see: the renormalized stress-energy tensor does not grow with the mass.

\subsection{Stress-energy tensor}

Due to the symmetries of de Sitter spacetime, the expression for the renormalized  stress-energy tensor can be obtained from its quantum trace as
\be \langle T^{\mu \nu} \rangle_r = \frac{1}{4} g^{\mu \nu} \langle T_{\rho}^{\rho} \rangle_r \ . \label{tensords} \ee
As analyzed in the last section, the formal expression for $\langle T_{\rho}^{\rho} \rangle$ contains UV divergences. From (\ref{trace}), (\ref{tracer}) and (\ref{tensords}), it is given by
\be  \langle T^{\mu \nu} \rangle =\frac{1}{4}g^{\mu\nu}\frac{-mH^{3}}{\pi^{2}}\int_{0}^{\infty}dz\left[\left(\left|h_{k}^{I}\right|^{2}-\left|h_{k}^{II}\right|^{2}\right)z^{2}\right] \ee
More specifically, it contains quadratic and logarithmic divergences, because for $z \rightarrow \infty$,
\be \left(\left|h_{k}^{I}\right|^{2}-\left|h_{k}^{II}\right|^{2}\right)z^{2}\overset{{\scriptstyle z\rightarrow\infty}}{\sim}z-\frac{1}{z}+\mathcal{O}\left(\frac{1}{z^{3}}\right) \ee
Therefore, in order to obtain the renormalized trace we subtract the corresponding adiabatic terms up to fourth order
\bea \label{Tdesitter}
\langle T^{\rho}_{\rho} \rangle_r & =&  \frac{-mH^{3}}{\pi^{2}} \times  \\
\int_{0}^{\infty} &  dzz^2 & \left( \left|h_{k}^{I}\right|^{2}-\left|h_{k}^{II}\right|^{2} - \left|g_{k}^{I\:\left(4\right)}\right|^{2} + \left|g_{k}^{II\:\left(4\right)}\right|^{2} \right) \nonumber 
 \ . \eea 
This integral is convergent and can be solved numerically. However, it can also be evaluated analytically by introducing an auxiliary regulator. This regulator has nothing to do with the regularization/renormalization process, which has already finished producing  the above finite expression for the renormalized trace.  The regulator that we are going to introduce now is a mere mathematic trick to evaluate analytically the finite integral (\ref{Tdesitter}). 

\begin{figure}
\begin{center}
\includegraphics[width=7.7cm]{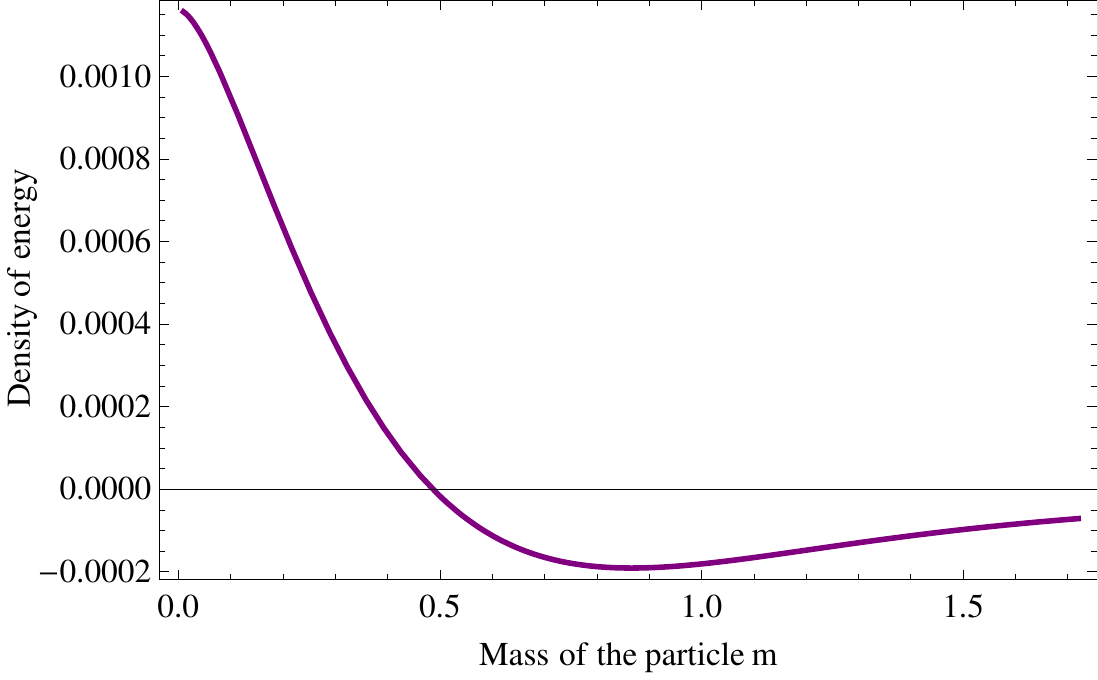} 
\end{center}
\caption{Density of energy $\rho = (1/4) \langle T^{00} \rangle$ as a function of the mass of the particle $m = \mu H$ for $H=1$.}
\label{figura3}
\end{figure}

First of all, the contribution of the exact modes to the integral (\ref{Tdesitter}) can be written as
\bea
\int_{0}^{\infty} & dz & z^2 \left(\left|h_{k}^{I}\right|^{2}-\left|h_{k}^{II}\right|^{2}\right)  = \int_{0}^{\infty}dz z^3 \left(\left|\widetilde{h}_{k}^{I}\right|^{2}-\left|\widetilde{h}_{k}^{II}\right|^{2}\right) \nonumber \\
& = & \lim_{\sigma\rightarrow0} -\frac{\partial^{2}}{\partial\sigma^{2}}\int_{0}^{\infty}dz z \cos\left(\sigma z\right) \left(\left|\widetilde{h}_{k}^{I}\right|^{2}-\left|\widetilde{h}_{k}^{II}\right|^{2}\right) \label{stensor1}
\eea
where we have defined $h_{k}^{I}=\sqrt{z}\widetilde{h}_{k}^{I}$ and  $h_{k}^{II}=\sqrt{z}\widetilde{h}_{k}^{II}$ in order to extract the dependence on powers of $z$ from $h_{k}$ and work with the regulator. The integration of (\ref{stensor1}) gives
\bea
 & \langle T^{\rho}_{\rho}\rangle &  =  \lim_{\sigma\rightarrow0}\left(\frac{H^{2}m^{2}}{\pi^{2}\sigma^{2}} - \frac{m^{2}\left(H^{2}+m^{2}\right)}{4\pi^{2}} \right. \times \nonumber \\
&&  \left. \frac{1+2\gamma+2\log\left(\frac{\sigma}{2}\right)+2\mathfrak{Re}\left[\psi\left(-2-i\frac{m}{H}\right)\right]}{4\pi^{2}} \right) \label{stres1}
\eea
which is divergent in the $\sigma \rightarrow 0$ limit. We can repeat the same procedure for the adiabatic subtraction integral
\bea
\int_{0}^{\infty} & dz & z^2 \left(\left|g_{k}^{I}\right|^{2}-\left|g_{k}^{II}\right|^{2}\right)  = \int_{0}^{\infty}dz z^3 \left(\left|\widetilde{g}_{k}^{I}\right|^{2}-\left|\widetilde{g}_{k}^{II}\right|^{2}\right) \nonumber \\
& = & \lim_{\sigma\rightarrow0} -\frac{\partial^{2}}{\partial\sigma^{2}}\int_{0}^{\infty}dz z \cos\left(\sigma z\right) \left(\left|\widetilde{g}_{k}^{I}\right|^{2}-\left|\widetilde{g}_{k}^{II}\right|^{2}\right) \label{stensor2}
\eea
and integrating (\ref{stensor2}), we obtain 
\bea \label{stress2}
&&  \langle T^{\rho}_{\rho} \rangle_{Ad} = \lim_{\sigma\rightarrow0}\left(\frac{H^{2}m^{2}}{\pi^{2}\sigma^{2}} -\frac{11H^{4}+190H^{2}m^{2}+60m^{4}}{240\pi^{2}}  \right. \nonumber \\
&& - \left. \frac{m^{2}\left(H^{2}+m^{2}\right)\left(2\gamma+2\log\left(\frac{\sigma}{2}\right)+2\log\left(\frac{m}{H}\right)\right)}{4\pi^{2}} \right) . \\ \nonumber 
\eea 
Equation (\ref{stress2}) is also divergent when $\sigma \rightarrow 0$. However, if (\ref{stress2}) is subtracted from (\ref{stres1}) the result  is finite in the $\sigma \rightarrow 0$ limit and gives the quantum trace. From it we can immediately obtain an analytic expression for the renormalized stress-energy tensor
\bea && \langle T^{\mu\nu} \rangle_r= \frac{1}{960 \pi^2} g^{\mu\nu} \left( 11 H^{4} + 130 H^{2} m^{2} \right. + \\
&& \left. 120m^{2}  ( H^{2}+m^{2} ) \left( \log\left(\frac{m}{H}\right)-\mathfrak{Re} \left[ \psi\left(-1 + i\frac{m}{H}\right) \right] \right) \right) \nonumber  \eea
where $\psi (z)$ is the digamma function. The function $ \langle T^{00} \rangle_r$ is shown in Fig. \ref{figura3} for $H = 1$. We observe that the energy density is bounded from above as a function of the mass.  In fact, for a large mass
$ \langle T^{\mu \nu} \rangle_r \sim 0 $, 
in sharp contrast with the behavior of the density of created particles obtained above. 

 \section{Conclusions}
 
 Since the original work \cite{parker-fulling74} introducing the systematics of the adiabatic regularization method for scalar fields, the extension of  the method to spin-$1/2$ fields has been lacking. We have developed here a satisfactory extension of the adiabatic method. 
The ansatz for the adiabatic
expansion for spin-$1/2$ field modes differs significantly from the WKB-type template that works for scalar
modes. We have tested the consistency of the
extended method by working out the conformal and axial anomalies for a Dirac field in a FLRW spacetime, in exact agreement with those obtained from other renormalization prescriptions. We have also given a detailed overview of the adiabatic prescription to analyze particle creation and renormalize expectation values of relevant physical observables. We have focused on the computation of particle creation in de Sitter spacetime.  Using the extended method we have been able to obtain an exact, analytical expression of the renormalized stress-energy
tensor for a Dirac field in de Sitter spacetime.

% \section*{Appendix I: Fourth adiabatic order for scalar fields}
 
%We provide here the explicit expression for $\omega^{(4)}$  in terms of the scale factor $a(t)$. 
% \bea & \omega^{(4)} = \frac{-1105 m^8 \dot a^4}{128 a^4 \omega^{11}} + \frac{221 (2 m^6 \dot a^4 + m^6 a \dot a^2 \ddot a )}{32 a^4 \omega^9} + \frac{138 m^4  \dot a^4}{32 a^4 \omega^7} + & \nonumber \\
% & \frac{ -300 \xi m^4 \dot a^4 -222 m^4 a \dot a^2 \ddot a -300 \xi m^4 a \dot a^2 \ddot a -19 m^4 a^2 \ddot a^2 -28 m^4 a^2 \dot a \dddot a}{32 a^4 \omega^7} + & \nonumber\\
%& \frac{-4 m^2 \dot a^4 + 72 \xi m^2 \dot a^4 - 15 m^2 a \dot a^2 \ddot a + 288 \xi m^2 a \dot a ^2 \ddot a + m^2 a^2 \ddot a^2}{16 a^4 \omega^5} + \nonumber \\
%& \frac{36 \xi m^2 a^2 \ddot a^2 + m^2 a^2 \dot a \ddot a + 60 \xi m^2 a^2 \dot a \dddot a + m^2 a^3 \ddddot a}{16 a^4 \omega^5} + & \nonumber\\ 
%& \frac{- \dot a^4 + 12 \xi \dot a^4 - 36 \xi^2 \dot a^4 + a \dot a^2 \ddot a + 6 \xi a \dot a^2 \ddot a - 72 \xi^2 a \dot a^2 \ddot a  + 2 a^2 \ddot a^2}{8 a^4 \omega^3} + & \nonumber \\
%& \frac{ - 6 \xi a^2 \ddot a^2 - 36 \xi^2 a^2 \ddot a^2 + 5 a^2 \dot a \dddot a  - 30 \xi a^2 \dot a \dddot a + a^3 \ddddot a - 6 \xi a^3 \ddddot a }{8 a^4 \omega^3} & \label{omega-ap} \nonumber .\\
%\eea

\section*{ Appendix: Fourth adiabatic order for spin-1/2 fields}
The first-, second-, and third order contributions to (\ref{nadiabatic}) are given in Sec. V. The fourth-order calculations are detailed in \cite{LMT}. The contributions with the simplifying condition, $F^{(4)} (m) = G^{(4)} (-m)$, are
{\small 
\bea  \omega^{\left(4\right)}  &=&\left[\dot{a}^{4}\left(-1105m^{6}+1348m^{4}\omega^{2}-377m^{2}\omega^{4}+12\omega^{6}\right)\right. \nonumber \\
  \;\;\;\;\;\;\;\;\;\;\;\;\;\;\,&+&2a\omega^{2}\dot{a}^{2}\ddot{a}\left(442m^{4}-389m^{2}\omega^{2}+52\omega^{4}\right) \nonumber \\
  &+&4a^{2}\omega^{4}\dot{a}\dddot{a}\left(-28m^{2}+15\omega^{2}\right)\nonumber \\
 &+& \left.4a^{2}\omega^{4}\left(\ddot{a}^{2}\left(-19m^{2}+8\omega^{2}\right)+2a\omega^{2}\ddddot{a}\right)\right]  \frac{m^{2}}{128a^{4}\omega^{11}} \nonumber \ ,\label{eq:imf4} 
\eea}
and
{\small
\begin{eqnarray}
F^{\left(4\right)} & = & \Biggl[\dot{a}^{4}\Biggl(9140 m^{7} -9040m^{6} \omega -10104m^{5} \omega^2 +10444m^{4} \omega^3  \nonumber \\
 &  &  + 2371 m^{3} \omega^4 -2664 m^2 \omega^5 - 48 m \omega^6 + 64 \omega^7 \Biggr)\nonumber \\
 &  & +8a\omega^{2}\dot{a}^{2}\ddot{a}\left(-914m^{5}+904m^{4}\omega+725m^{3}\omega^{2}-749m^{2}\omega^{3} \right. \nonumber \\
 &  & \left.-76m\omega^{4} +88\omega^{5}\right) +32a^{2}\omega^{4}\dot{a}\dddot{a}\left(28m^{3}-28m^{2}\omega-13m\omega^{2} \right. \nonumber \\
 &  & \left. +14\omega^{3}\right) +4a^{2}\omega^{4}\left(4\left(m-\omega\right)\left(41m^{2}+\left(m-16\omega\right)\omega\right)\ddot{a}^{2} \right. \nonumber \\
 & & \left. +16a\omega^{2}\ddddot{a}\left(-m+\omega\right)\right)\Biggr] \frac{m}{2048a^{4}\omega^{12}}\:.\nonumber 
\end{eqnarray}
}
\\

{\it Acknowledgments:}  J. N-S. would like to thank I. Agullo, G. Olmo and L. Parker for very useful discussions. 
This work is supported  by the  Spanish Grant No. FIS2011-29813-C02-02 and  the Consolider Program CPANPHY-1205388.

\end{document}